\documentclass[prc,amsmath,twocolumn,showpacs,superscriptaddress]{revtex4}
\usepackage[dvips]{graphicx}
\usepackage{dcolumn}

\newcommand{\fig}[1]{Fig.~\ref{#1}}
\newcommand{\figs}[2]{Figs.~\ref{#1} and \ref{#2}}

\begin{document}

\title{$^7$Be breakup on heavy and light targets}

\author{N.~C.~Summers}
 \email{summers@nscl.msu.edu}
 \affiliation{National Superconducting Cyclotron Laboratory, 
Michigan State University, East Lansing, Michigan 48824}
\author{F.~M.~Nunes}
 \affiliation{National Superconducting Cyclotron Laboratory, 
Michigan State University, East Lansing, Michigan 48824}
 \affiliation{Department of Physics and Astronomy, 
Michigan State University, East Lansing, Michigan 48824}
\date{19 April 2004}

\begin{abstract}
In this paper we present all-order quantum mechanical 
calculations of $^7$Be breakup on heavy ($^{208}$Pb) and light ($^{12}$C) targets.
We examine the issues concerning the extraction of the astrophysical $S$-factor $S_{34}(0)$ from the breakup data.
We discuss the interplay between Coulomb and nuclear breakup, 
and the importance of higher-order couplings on the cross section.
We show that nuclear and Coulomb contributions are not separable 
using the standard angular selection criterion 
as nuclear breakup remains large for small scattering angles,
even for the heavy target. 
However, by selecting an upper limit on the relative energy between the final fragments, the contribution from the nuclear breakup can be significantly reduced such that Coulomb breakup is the main reaction mechanism.
We show that the extraction of the asymptotic normalization coefficient may require more careful consideration of the nuclear interior than previously used.

\end{abstract}


\maketitle

At present, the capture rate $^3$He($\alpha$,$\gamma$)$^7$Be is now more
uncertain than $^7$Be($p$,$\gamma$)$^8$B, both belonging to the $pp$ chain and with connections to the solar neutrinos \cite{adelberger-98}.
Also, the $^3$He($\alpha$,$\gamma$)$^7$Be reaction is the only important production channel for $^7$Li in big-bang nucleosynthesis \cite{cyburt-04}.
Reference~\cite{adelberger-98} recommends a value of
$S_{34}(0) = 0.53 \pm 0.05$ keV b.
At the low energies of astrophysical relevance, capture rates are exceptionally hard to measure directly and thus rely on extrapolations.
Recent theoretical studies claim that the extrapolation of $S_{34}(0)$ can be significantly more uncertain \cite{csoto-00}. 

Consequently, measurements with alternative methods have been considered.
There are two main methods in use:
i) The Coulomb dissociation method proposes the measurement of 
radiative capture rates from breakup data on heavy targets \cite{baur-86},
ii) The asymptotic normalization coefficient (ANC) method 
proposes the measurement of the capture rate 
from transfer reactions \cite{xu-94} or breakup reactions \cite{trache-01}.
Both have been applied successfully to other reactions \cite{davids-01,trache-01}.

The uncertainties on the direct capture measurements for
the $^3$He($\alpha$,$\gamma$)$^7$Be reaction have motivated two 
experiments: one at the NSCL, measures the $^7$Be breakup
on $^{208}$Pb at 100 MeV/nucleon, and the other, at the Cyclotron Lab
at Texas A \& M, will use a $^7$Be 25 MeV/nucleon beam on a
$^{12}$C target. It is therefore timely to consider the 
reaction mechanisms involved in extracting $S_{34}(0)$ from these reactions.

The cross sections for measuring the breakup of $A \rightarrow c+x$ via the
Coulomb field of a heavy target $T$ are much larger than the low energy 
direct capture cross sections.
From the Coulomb dissociation data involving low relative breakup 
energies between the $c+x$ fragments, 
one can extract the inverse reaction $c+x \rightarrow A$ 
at astrophysical energies \cite{baur-86,motobayashi-94}.
There are three main breakup mechanisms which can complicate this relation:
i) nuclear breakup, present whenever the projectile gets close to the target,
ii) E2 transitions, may be negligible in radiative capture, 
but are typically significant in Coulomb dissociation \cite{davids-98},
iii) final state interactions (continuum-continuum (CC) couplings) that 
distort the final energy spectrum of the emitted fragments \cite{gazes-92}.
All of these entangle the information on the capture reaction.
In this paper we examine these various issues for the
particular case of the $^7$Be breakup,
making use of the continuum discretized coupled channels (CDCC) method, 
reviewed in Ref.~\cite{cdcc-theory}.
Within CDCC, nuclear and Coulomb are consistently included and
the contribution from CC couplings can be explicitly explored.

There have been several applications of the Coulomb dissociation
method to other astrophysically important breakup reactions.
The nuclear and Coulomb interplay has been studied for the 
$^8$B $\to$ $^7$Be+$p$ breakup reaction 
\cite{nunes-98,nunes-99,thompson-01,esbensen-99}.
Here, Coulomb dissociation dominates for small scattering angles, 
which can be related semi-classically to large impact parameters 
outside the range of the nuclear force.
Studies of $^7$Li and $^6$Li resonant breakup have suggested that 
for these nuclei, at high energies, the Coulomb and nuclear breakup 
can be well separated into angular regions, and the nuclear effects 
are small at higher energies and forward angles \cite{shyam-91}.
Conversely, earlier studies of $^7$Li breakup have shown that 
experimental data differ significantly from Coulomb-only calculations 
at forward angles which indicates that the nuclear forces are important 
even at forward angles \cite{shotter-88a,shotter-88b,shotter-89}.
In addition, coupled channels calculations of sequential breakup 
via the $3^+$ resonance in $^6$Li have been performed, and again nuclear 
effects were found to be important even at forward angles \cite{hirabayashi-92}.

From these studies of similar reactions 
it is not possible to derive the importance of nuclear effects for the 
$^7$Be breakup reactions of interest here.
We therefore perform fully-quantum mechanical calculations,
in the CDCC framework using the coupled-channels code \textsc{fresco} \cite{fresco},
of $^7$Be breakup on $^{208}$Pb and $^{12}$C targets at energies of
100 and 25 MeV/nucleon respectively.

The breakup of $^7$Be into $\alpha$+$^3$He by the interaction with a target
consists of a three-body (two-body projectile + target) problem.
In the CDCC method, the breakup of the $^7$Be projectile is treated as
an excitation into the $\alpha$+$^3$He continuum, discretized into $\mathcal N$ bins.
For $^7$Be, with two bound states below the $\alpha$+$^3$He breakup threshold, this requires a $\mathcal N+2$ coupled channels problem.
The physical inputs are then: 
an $\alpha$+$^3$He potential which binds the $^7$Be, and for 
each cluster, an interaction with the target, which 
can be complex to account for loss of flux to unaccounted channels.

The $\alpha$+$^3$He potential used here is that of Buck {\it et al.} \cite{buck-85}.
This potential consists of a central and spin orbit terms of Gaussian form.
The parameters were fixed using the binding energies of the bound states and the positions of the resonances for $^7$Be and $^7$Li, the $^7$Li charge radius, quadrupole and octupole moments, and B(E2:$3/2^-\to1/2^-$).
We increased slightly the spin-orbit potential depth to fit exactly the $^7$Be binding energies for the $2p_{3/2}$ ground state ($S_\alpha$=1.587 MeV) and $2p_{1/2}$ first excited state ($S_\alpha$=1.158 MeV), since these fix the correct asymptotics of the bound state wavefuntions.
The bound states have a node because the lower orbitals are excluded due to Pauli blocking.
This potential produced the resonances in the $f$-waves at approximately the correct energies.
In Ref.~\cite{buck-85}, the potential depth was adjusted to fit the $s$-wave phase shifts, which also gave small $d$-wave phase shifts as required.
Therefore, we use a parity dependent potential, with a depth of 
$V_{\mathrm{odd}}$ = 83.77 MeV for the $p$- and $f$-waves, and 
$V_{\mathrm{even}}$ = 66.10 MeV for the $s$- and $d$-waves. 
The spin orbit depth is $V_{ls}$ = 3.8 MeV, and the range for all 
three potentials is $R_{\mathrm{odd}}=R_{\mathrm{even}}=R_{ls}=2.52$ fm. 
The Coulomb radius is $R_C$=3.095 fm.

The calculated B(E2:$3/2^-\to1/2^-$) for this interaction is 19.0 e$^2$fm$^4$, consistent with Ref.~\cite{mertelmeier-86}.
The Buck model produces a rms charge radius for $^7$Be of 2.62 fm, 
in reasonable agreement with the experimentally 
determined charge radius of 2.52 $\pm$ 0.03 fm \cite{tanihata-85}.

Optical potentials are important ingredients in our calculations.
The cluster-target potentials are fixed by elastic scattering data at the energy carried by that cluster in the projectile.
We require $^3$He+$^{208}$Pb at 300 MeV and $\alpha$+$^{208}$Pb at 400 MeV.
The nearest available energy for elastic data was at 217 MeV for $^3$He+$^{208}$Pb \cite{he3pb208e217} and 340 MeV for $\alpha$+$^{208}$Pb \cite{alphapb208}.
We require $^3$He+$^{12}$C at 75 MeV and $\alpha$+$^{12}$C at 100 MeV.
We use the potential for $^3$He+$^{12}$C at 72 MeV from Ref.~\cite{demyanova-92} but neglect the spin-orbit term, 
and the potential for $\alpha$+$^{12}$C at 104 MeV from Ref.~\cite{hauser-69}.
Different sets of potential parameters for the fits to the elastic data are reported in the literature.
The sensitivity of the breakup cross section to the various potentials was found to not significantly effect the breakup cross section
(less than 2\% at $0^\circ$ for both targets).

\begin{figure}
\includegraphics[width=8cm]{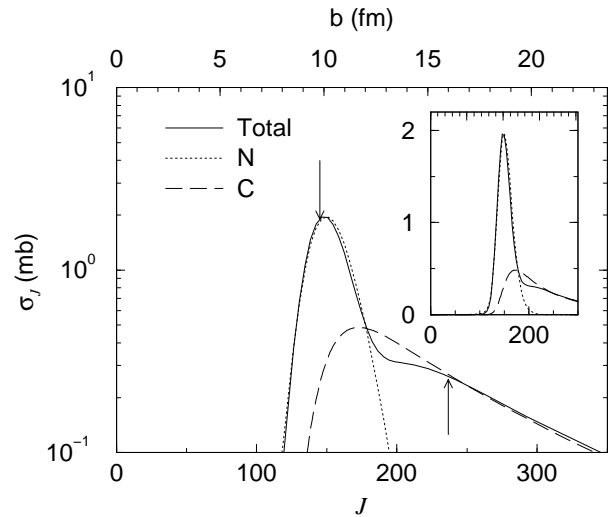}
\caption{\label{FIG:pb208-sigj} $J$-distribution of the cross section for the $^{208}$Pb target. The total breakup cross section is shown by the solid line while the broken lines give the nuclear (dotted) and Coulomb (dashed) contributions. The top scale relates the angular momentum to the impact parameter via the semi-classical relation $J=Kb$, where $K=15$ fm$^{-1}$. The insert is the same plot on a linear scale. The arrows are discussed in the text.}
\end{figure}
\begin{figure}
\includegraphics[width=8cm]{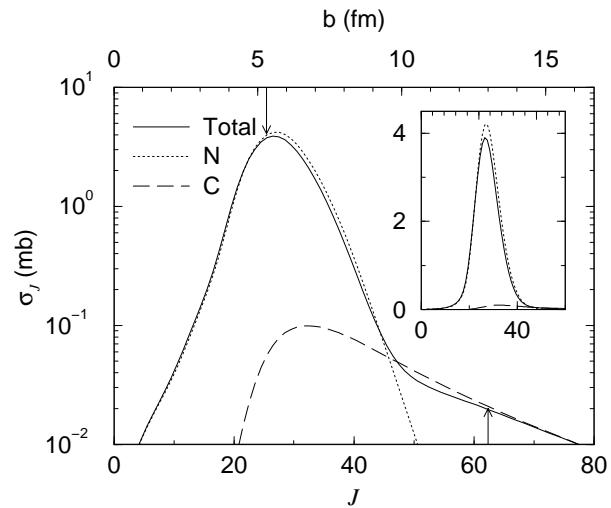}
\caption{\label{FIG:c12-sigj} $J$-distribution of the cross section for the $^{12}$C target. The lines have the same meaning as \fig{FIG:pb208-sigj}, and the impact parameter scale on the top axis uses $K=4.8$ fm$^{-1}$.}
\end{figure}

In the present calculation, the $^3$He+$\alpha$ continuum includes partial 
waves for $\ell\le3$ and each $j^\pi$ discretized into 10 bins up to 20 MeV, equally spaced in momentum.
Due to the sharp resonance, the $f_{7/2}$ partial wave has a resonance 
bin with the width of the resonance, and two bins below the resonance bin.
Above the resonance, we include 8 bins up to 20 MeV for the lead target 
and 9 bins up to 30 MeV for the carbon target.
Convergence of this discretization of the continuum was checked by doubling the number of bins, which only affected the total breakup cross section by 2\%.
The continuum bins are integrated out to a maximum radius $r_\mathrm{bin}=50$ fm.
Potential multipoles are included for $q\le2$ as octupole transitions are negligible.
Using interpolation, partial waves up to $L=14000 \hbar$ are included for the lead target and up to $L=2000 \hbar$ for the carbon target.
The coupled equations are solved using $R_\mathrm{max}=1000$ fm.
We use non-relativistic kinematics which, at 100 MeV/nucleon on the lead target, will introduce a 3\% error in the momentum.

In \fig{FIG:pb208-sigj} (\fig{FIG:c12-sigj}) we show the $J$-distribution of the breakup cross section for the $^{208}$Pb ($^{12}$C) target.
The impact parameters that correspond to each partial wave, using the 
semi-classical relation $J=Kb$, are shown across the top scale.
The total breakup cross section (solid line) is shown along with the nuclear breakup (dotted) and Coulomb breakup (dashed).
The sum of the radii for the projectile and target is marked on the figures by the down-pointing arrow.

Expectedly, we see that for both light and heavy targets, the nuclear 
breakup is the dominant process for the lower partial waves and Coulomb 
breakup dominates the higher partial waves.
The Coulomb breakup, suppressed at low partial waves due to the nuclear absorption, peaks around the sum of the projectile and target radii, then falls off slowly with increasing $J$.
The nuclear breakup is small for the low partial waves since the imaginary 
part of the nuclear potential is removing flux from elastic breakup 
to other channels.
It peaks sharply around impact parameters corresponding to surface collisions 
of the two nuclei, then falls off rapidly.

To extract the ANC from which $S_{34}$(0) can be determined, breakup data from a range of targets can be used.
The fundamental requirement is peripherality.
A simple sum of radii would imply that there should be no contribution to the breakup
from impact parameters below 9.8 (5.3) fm for the lead (carbon) target.
The plots of \figs{FIG:pb208-sigj}{FIG:c12-sigj} suggest that this may not be the case.
The breakup cross section for impact parameters below the sum of radii is 28\% (16\%) of the total breakup cross section for the lead (carbon) target.

\begin{figure}[t]
\includegraphics[width=8cm]{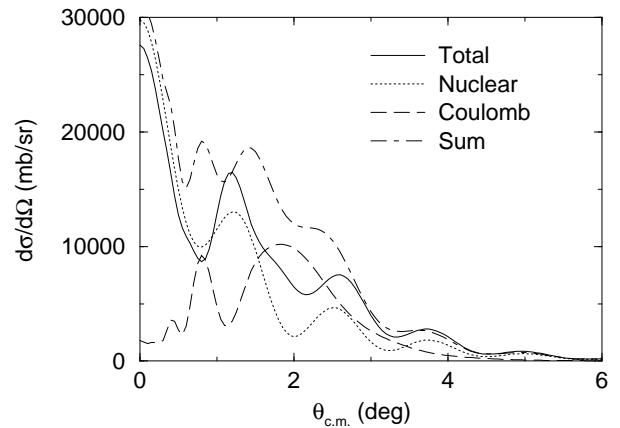}
\caption{\label{FIG:pb208-n+c} Angular distribution of cross sections for $^7$Be elastic breakup on $^{208}$Pb at 100 MeV/nucleon. The CDCC (solid) calculation is broken down into nuclear (dotted) and Coulomb (dashed) contributions. The sum (dot-dashed) represents the incoherent sum of the nuclear and Coulomb contribtutions, while the solid line is the coherent sum.}
\end{figure}
\begin{figure}[t]
\includegraphics[width=8cm]{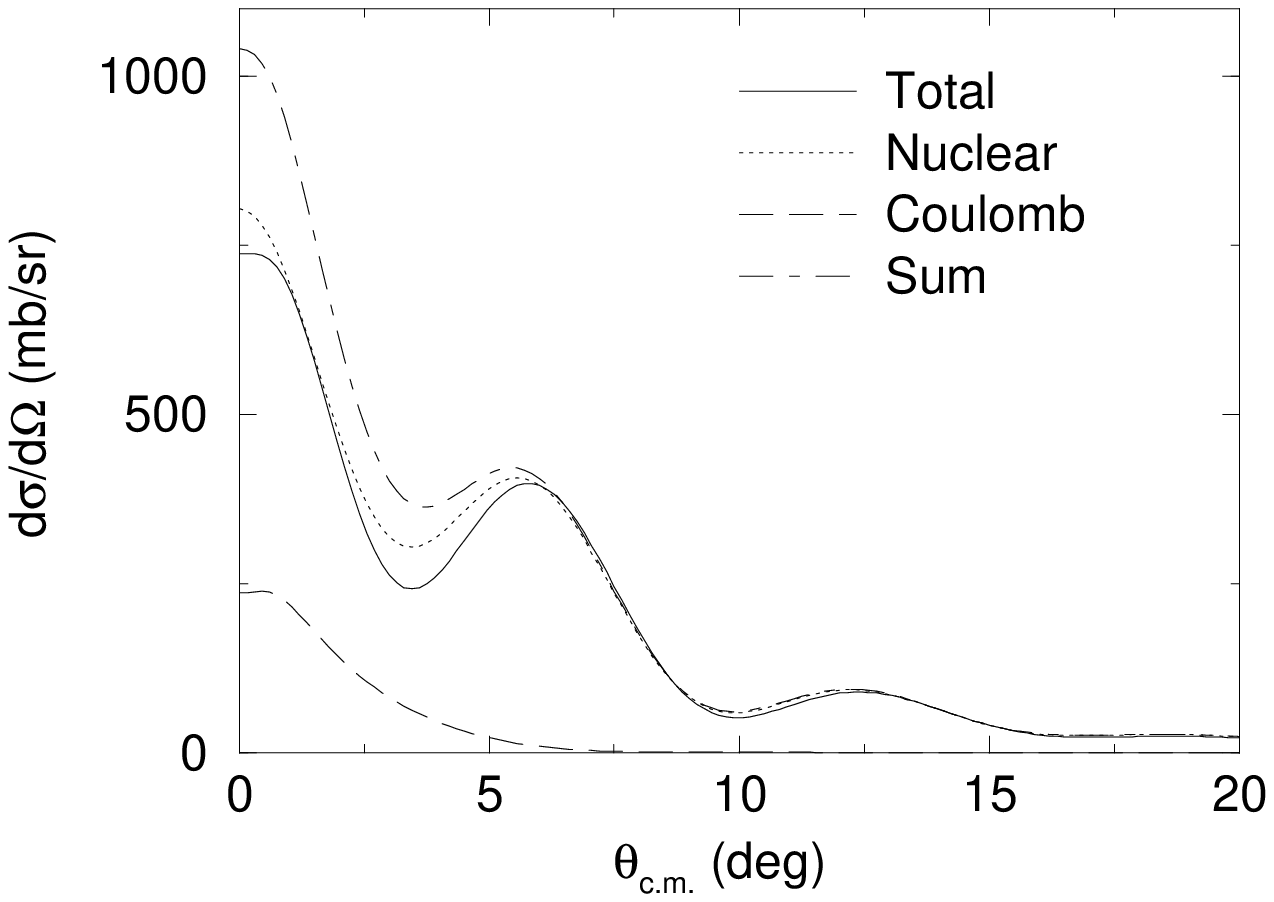}
\caption{\label{FIG:c12-n+c} Angular distribution of cross sections for $^7$Be elastic breakup on $^{12}$C at 25 MeV/nucleon. The lines have the same meaning as \fig{FIG:pb208-n+c}.}
\end{figure}
\begin{figure}[t]
\includegraphics[width=8cm]{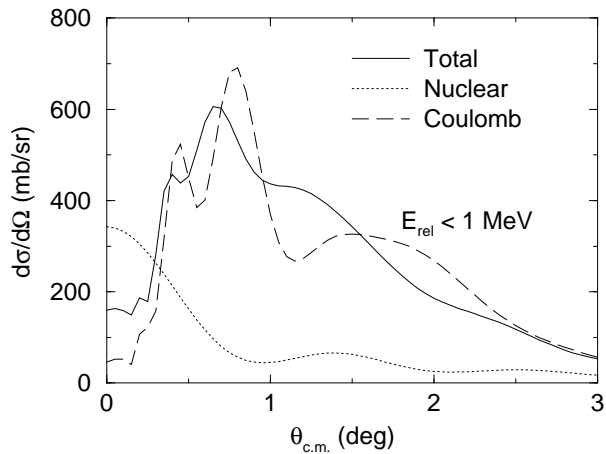}
\caption{\label{FIG:pb208-n+c-e1} Angular distribution of the breakup cross section in \fig{FIG:pb208-n+c} which includes only the lowest two energy bins from each $j^\pi$ set. This gives the relative energy between the $\alpha$ and $^3$He fragments an upper limit of approximately 1 MeV.}
\end{figure}
\begin{figure}[t]
\includegraphics[width=8cm]{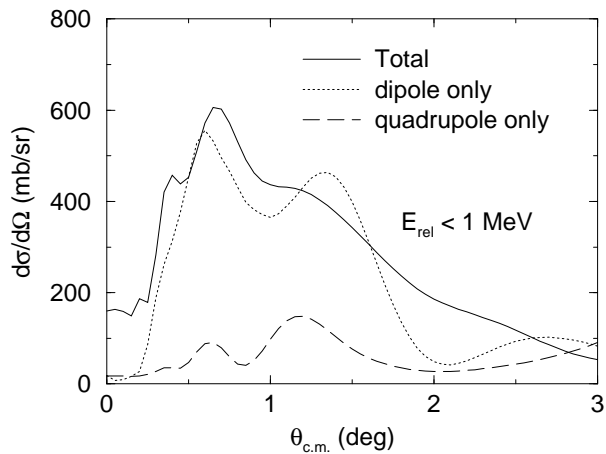}
\caption{\label{FIG:pb208-q1q2-e1} Angular distribution of the breakup cross section showing the dipole and quadrupole contributions. Both nuclear and Coulomb are included. As with \fig{FIG:pb208-n+c-e1}, this includes only the lowest two energy bins from each $j^\pi$ set. This gives the relative energy between the $\alpha$ and $^3$He fragments an upper limit of approximately 1 MeV.}
\end{figure}

The slower fall off for Coulomb breakup means that the Coulomb dominates 
beyond a certain partial wave, where the nuclear breakup is negligible and 
interference effects are small.
The impact parameter, beyond which nuclear effects can be considered negligible (up-pointing arrow on figures), 
is around 16 (13) fm the lead (carbon) target, values much larger than the sum of the
projectile and target radii.
In the Coulomb dissociation method, data are typically taken at 
forward angles since then it is assumed to be nuclear free.
For pure Rutherford trajectories, there is a relationship between
the impact parameter and the scattering angle: for  
$^7$Be+$^{208}$Pb at 100 MeV/nucleon, an impact parameter of 16 fm corresponds 
to a center-of-mass scattering angle of 2.5$^\circ$;
for $^7$Be+$^{12}$C at 25 MeV/nucleon, 
13 fm corresponds to an angle of 1.4$^\circ$.
However, the determination of these cutoff angles is rather simplistic.
One should note that the Coulomb breakup cross section here differs 
significantly from  the semi-classical Coulomb dissociation cross section \cite{cdxs}.
This is mainly due to the finite size of the projectile which has been previously
pointed out in the CDCC calculations of Ref.~\cite{nunes-98}.

\begin{figure}[t]
\includegraphics[width=8cm]{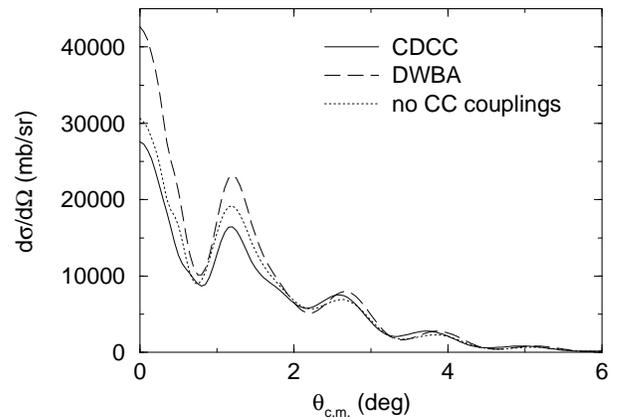}
\caption{\label{FIG:couplings-pb} Angular distribution of the cross section for $^7$Be breakup on $^{208}$Pb at 100 MeV/nucleon. The different calculations of the cross section are: CDCC (solid), DWBA (dashed), and a CDCC calculation without continuum-continuum (CC) couplings.}
\end{figure}
\begin{figure}[t]
\includegraphics[width=8cm]{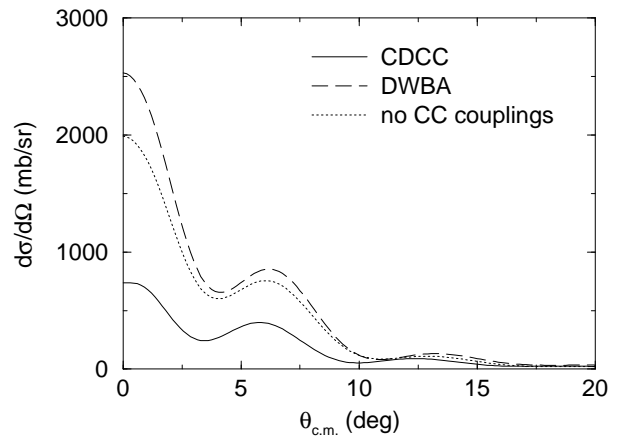}
\caption{\label{FIG:couplings-c} Angular distribution of the cross section for $^7$Be breakup on $^{12}$C at 25 MeV/nucleon. The lines have the same meaning as \fig{FIG:couplings-pb}.}
\end{figure}

We can directly examine the angular distribution of the breakup cross section 
on the lead (carbon) target, shown in \fig{FIG:pb208-n+c} (\fig{FIG:c12-n+c}).
The nuclear (dotted) and Coulomb (dashed) contributions to 
breakup are plotted along with the coherent sum (solid).
The interference between nuclear and Coulomb breakup is shown by the 
difference between the incoherent sum (dot-dashed) and the coherent sum (solid).
We see that for both targets the nuclear breakup is the 
dominant process, having a diffractive nature which peaks at zero degrees.
Contrary to expectations, we see that the nuclear and Coulomb breakup cannot be separated into angular regions and both have to be considered.

However for the heavy target, the nuclear breakup contribution can be reduced by imposing an upper cut on the relative energy between the $\alpha$ and $^3$He fragments.
To show the effect of this energy cut on the angular distribution we sum up the angular cross section from the lowest two energy bins from each $j^\pi$ set.
This restricts the maximum final state relative energy to approximately 1 MeV (0.846 MeV for all $j^\pi$ except the $f_{7/2}$ which has a maximum energy of 1.268 MeV). 
The angular distribution of the breakup cross section with this energy cut is shown in \fig{FIG:pb208-n+c-e1}.
We now see that Coulomb breakup is dominant except for the extreme forward angles (less than 0.5$^\circ$).
To give an estimate of the E2 contribution to the breakup cross section,
we plot the dipole and quadrupole contributions to the breakup (solid line in \fig{FIG:pb208-q1q2-e1}, which includes nuclear and Coulomb breakup).
Once again the maximum relative energy has been restricted to approximately 1 MeV so there is a Coulomb dominated region up to $\theta_{\mathrm{c.m.}}=3^\circ$, as shown in \fig{FIG:pb208-n+c-e1}.
We see that in this region, dipole transitions dominate.

For lighter targets, the nuclear contribution to breakup is usually large
and Coulomb breakup is considered small.
We see from \fig{FIG:c12-n+c} that this is the case above 6$^\circ$.
Below 6$^\circ$ the coherent sum of the nuclear+Coulomb breakup (solid) is very 
similar to that of nuclear (dotted) breakup alone, although this does 
not mean that Coulomb effects are negligible.
Below 6$^\circ$, the nuclear and Coulomb breakup contributions interfere,
through strong couplings in the continuum,
to give a coherent sum which is similar to that of nuclear alone.

The effect of couplings on the angular distribution is shown in 
\fig{FIG:couplings-pb} (\fig{FIG:couplings-c}) for the lead (carbon) target.
The solid line is the full CDCC calculation which includes all 
couplings to all orders.
The DWBA calculation (dashed) only includes first order couplings 
between ground and excited states. The dotted line shows a subset 
of the CDCC calculation which neglects the continuum-continuum (CC) couplings.
We see that for both targets the couplings reduce the cross section 
significantly, less so for the heavier target, where the experiment 
was performed at a higher beam energy.
When extracting astrophysical quantities from this breakup data, it
is important to go beyond first-order reaction theories.

In conclusion, we have shown that for these breakup measurements, contributions from the interior are significant.
Selected cuts on the data need to be considered for the extraction of the ANC, which requires a peripheral collision.
In addition, the relationship between the ANC and the astrophysical $S$-factor relies on a first order DWBA approach, but we have shown that the DWBA cross section dramatically overestimates the breakup cross section.
It is not clear whether this method will be helpful in improving $S_{34}(0)$.
We have shown that by performing $^7$Be breakup experiments on 
heavy targets, it is not possible to completely eliminate 
nuclear effects through selecting $\theta_{\mathrm{c.m.}}$ smaller than a critical value.
However, by selecting an upper limit on the relative energy between the final fragments, the contribution from the nuclear breakup can be significantly reduced such that E1 Coulomb breakup is the main reaction mechanism.
The extraction of $S_{34}(0)$ using the Coulomb dissociation method may be promising, but as yet, the uncertainties due to the nuclear and E2 contributions, along with the experimental uncertainties in the measurement, makes it unclear whether this method will be able to improve on the radiative capture measurements.

The authors acknowledge useful discussions with 
P.\ G.\ Hansen, Sam M.\ Austin and C.\ A.\ Bertulani.
This work is supported by NSCL, Michigan State University.

%\newpage
\bibliography{ref,be7}

\end{document}